\documentclass[oldversion]{aa} 
\usepackage{graphicx}
\usepackage{txfonts}
\begin{document} 

\input psfig.tex
\titlerunning{JKCS041, a galaxy cluster at $z\sim1.9$}
\authorrunning{Andreon et al.}
\title{JKCS\,041: a colour--detected galaxy
cluster at $\mathbf{z_{phot}\sim1.9}$ with deep potential well as confirmed by X-ray
data}
\author{S. Andreon
\inst{1},
B. Maughan
\inst{2}, 
G. Trinchieri,
\inst{1} \and
J. Kurk
\inst{3}
}
\institute{
INAF--Osservatorio Astronomico di Brera, via Brera 28, 20121, Milano, Italy \\
Department of Physics, University of Bristol, Tyndall Ave, Bristol BS8 1TL, UK\\
Max-Planck-Institut fur Astronomie, Konigstuhl, 17 D-69117, Heidelberg, Germany\\
\email{stefano.andreon@brera.inaf.it}
}
\date{Received date; accepted date}
\abstract{
We report the discovery of JKCS\,041, a massive near-infrared selected
cluster of galaxies at $z_{phot} \sim1.9$. The cluster was originally discovered using a modified red-sequence
method and was also detected in follow-up Chandra
data as an extended X-ray source. Optical and near-infrared imaging data alone allow us
to show that the detection of JKCS\,041 is secure, even in absence of the X-ray data.
We investigate the possibility that JKCS\,041 is not a galaxy cluster at $z\sim1.9$, 
and find other explanations unlikely. The X-ray detection and statistical arguments rule out the hypothesis
that JKCS\,041 is actually a blend of groups along the line of sight, and we find that the X-ray emitting gas is 
too hot and dense to be a filament projected along the 
line of sight. The absence of a central radio source and the extent and
morphology of the X-ray emission argue against the possibility that
the X-ray emission comes from inverse Compton scattering of CMB photons
by a radio plasma. The cluster has
an X-ray core radius of $36.6^{+8.3}_{-7.6}$ arcsec (about 300 kpc), an
X-ray temperature of $7.4^{+5.3}_{-3.3}$ keV,  
a bolometric X-ray luminosity within
$R_{500}$ of $(7.6\pm0.5)\times 10^{44}$ erg s$^{-1}$, and an estimated mass of 
$M_{500}=2.9^{+3.8}_{-2.4}\times 10^{14}M_\odot$, the last
derived under the usual (and strong) assumptions. 
The cluster is composed of $16.4\pm6.3$ galaxies within 1.5 arcmin (750
kpc) brighter than $K\sim 20.7$ mag.
The high redshift of JKCS\,041 is determined
from the detection colour, from the detection of
the cluster in a galaxy sample formed by $z_{phot}>1.6$ galaxies 
and from a photometric redshift based on 11-band spectral energy distribution fitting. 
By means of the latter we find the cluster redshift to be
$1.84<z<2.12$ at 68 \% confidence.
Therefore, JKCS\,041 is a cluster of galaxies at $z_{phot} \sim 1.9$ with a deep
potential well, making it the most distant cluster 
with extended X-ray emission known. 
}


\maketitle

\section{Introduction}

Clusters of galaxies are known to harbour red galaxies out to the highest
redshifts explored thus far, $z=1.45$ (Stanford et al. 2006). They owe their
colour mostly to their old stellar populations:   
their luminosity function is passively
evolving (de Propris et al. 1999, Andreon 2006a, Andreon et al. 2008) 
and their
colour-magnitude relation evolves in slope, intercept
and scatter as expected for a passive evolving population (e.g. Stanford et al. 1999; 
Kodama et al. 1998). The presence of red galaxies characterises
clusters irrespectively of the way they are selected:  X-ray selected
clusters
show a red sequence (see, e.g. Andreon et al. 2004a, 2005, for XMM-LSS
cluster samples at $0.3<z<1.2$, Ebeling et al. 2007 for MACS, whose
red sequence is reported in Stott et al. 2007 and Andreon 2008, see also
Stanford et al. 2005, Lidman et al. 2008, Mullis et al. 2005 
for some other individual clusters). The first Sunyaev-Zeldovich
selected clusters (Staniszewski et al. 2008) have been confirmed 
thanks to their red sequence, and their photometric
redshifts were derived from its colour. The same is often true 
for shear-selected clusters (e.g. Wittman et al. 2006), starting with 
the first such example (Wittman et al. 2001).
 
It is now established that bright red galaxies exist at
the appropriate frequency in high redshift clusters and the discussion
has now shifted to the faint end of the red population. 
Current studies aim at
establishing whether their abundance evolves with lookback
time, with proponents divide into opposite camps (e.g 
Andreon 2008; Lidman et al. 2008; Crawford et al. 2008, Tanaka et al. 2008
vs Gilbank \& Balogh 2008, Stott et al. 2007). However, 
these galaxies are too faint for the purpose of discovering 
clusters, and thus such discussion is not relevant here.

On the contrary, bright red galaxies are a minority in the field
population. Consequently, selections based on galaxy colour
enhance the contrast of the large red galaxy population in clusters relative to
bluer field galaxies.

The Balmer break is a conspicuous characteristic feature of galaxy
spectra. Red colours are observed when a galaxy has a prominent 
Balmer break and is at the appropriate redshift for the filter
pair used. In some circumstances a galaxy may look red,
when another spectral break (e.g. Lyman break) 
falls between the filter pair, but these galaxies are numerically few in actual samples
and are not clustered as strongly as the galaxies in clusters. Therefore,
their clustering cannot be mistaken as a cluster 
detection. Indeed, if they were found to be so strongly clustered 
it would be an interesting discovery in itself.
The colour index thus acts as a (digital) filter and
removes galaxies with a spectral energy distribution
inconsistent with the expected one. One can then select
against objects at either a different redshift or having
unwanted colours, making the colour selection a 
very effective way of detecting clusters.
Red-sequence-like algorithms, pioneered by Gladders \& Yee (2000)
for detecting clusters rely precisely
on identifying a spatially localised overdensity of
galaxies with a pronounced Balmer break.

Several implementations of red-sequence-like algorithms have been used to
detect clusters (e.g. Gladders \& Yee 2000; Goto et al. 2002, Koester et al. 2007),
each one characterised by different assumptions on the cluster model (i.e. 
on the expected properties of the ``true" cluster). For example, some models require clusters to follow the
Navarro, Frenk \& White (1997) radial profile with a predicted radius scale. Alternatively the scale (or form of the profile) can be left free (or only weakly bounded). To
enlarge the redshift baseline in such cluster surveys, one is simply required to change filters in order to ``follow"
the Balmer break to higher and higher  redshifts. 
One version of these algorithms has been applied in a series of papers by Andreon
et al.' to cover the largest redshift range sampled in one study. In
Andreon (2003) the technique was applied to the nearby ($z<0.3$) universe using SDSS
$g$ and $r$ filters.  With the $R$ and $z'$ filters  the medium-distant
universe ($0.3 \la z < 1.1$) was probed (Andreon et al. 2004a,b, 2005,
2008, note that the same bands are used by the
red sequence survey, Gladders \& Yee 2005). To sample the $z\approx 1.2$
universe, infrared bands must be used for the ``red" filter, because the Balmer
break moves into the $z'$ band at $z \sim 1.2$. This is shown by Andreon et al.
(2005), who were able to detect XLSSC 046 (called ``cluster h" in that paper) at
$z=1.22$ using $z'$ and $J$, but we missed it in $R$ and $z'$.  XLSSC  046 was
later spectroscopically confirmed in Bremer et
al. (2006). 

Cluster detection by red sequence-like method is observationally 
cheap: it is possible
to image large ($\ga 1$ deg$^2$ per exposure) sky regions 
at sufficient depth to detect clusters up
to $z=1$, with short ($\la 2$ ks) exposure times on 4m ground-based 
telescopes.  All-sky X-ray surveys such as the RASS also provide 
efficient cluster detections, 
although they do require more expensive space-based observations.
For distant clusters, 
currently operational X-ray observatories are less efficient than red-sequence methods due to the longer
exposure times required (e.g.
$> 10$ ks) and smaller fields of view
(1/9 deg$^2$ at most). 
As a result, red-sequence-like studies are able to follow the cluster mass function down to lower mass limits than is possible with X-ray observations
(e.g. Andreon et al. 2005).   
On the other hand, X-ray detected clusters are less affected
by confusion issues, i.e. by the possibility that the detected structure is
a line-of-sight blend of smaller structures. 
Cluster detection by gravitational shear, meanwhile, requires deeper optical data than
red-sequence-like algorithms. This is because shear studies require the measuement of subtler quantities (small distortions in shape) of fainter galaxies. Furthermore, detection
by shear is
affected by confusion (we will return to this issue in sec. 3.6), and thus 
may be of limited use for
detecting samples of clusters. However, shear measurements offer a more direct
probe of cluster mass, provided the shear detection has a sufficient
signal to noise to perform the measurement in survey data 
(although this is currently not common, see e.g. Schirmer et al. 2007).

In this paper, we present the detection of the very distant,
$z_{phot}=1.9$ cluster JKCS\,041\footnote{Cluster names are acronyms indicating
the filter pair used for the detection (gr,Rz,JK) followed by the string
CS, for colour selected, followed by the order
number in the catalogue. These names are IAU-compliant as the acronyms are registered.} obtained by adopting redder ($J$ and $K$) filters 
to follow the Balmer break
to even higher redshifts.  Since the Balmer break enters
the $J$ filter at $z\sim1.9$, 
$J-K>2.3$ mag (in the Vega system) is a very effective 
criterion (Saracco et al. 2001,
Franx et al. 2003) for selecting $z\ga 2$ galaxies. 
The spectroscopic study by Reddy et al. (2005), by 
Papovich et al. (2006) and by Kriek et al. (2008), all confirme
that the $J-K>2.3$ mag criterion selects mainly galaxies at $z\ga2$.

The paper is organised as follow:
Section 2 presents the original cluster discovery. In section 3 we reinforce 
the cluster detection using different methods and we determined the cluster
photometric redshift. In that section, we also show that JKCS\,041 is
not a blend of two (or more) groups along the line of sight. JKCS\,041 is
also X-ray detected in follow-up Chandra observations. Section 4 describes
these data, confirming the cluster detection, and presents our measurements of basic cluster 
properties (core radius, luminosity, temperature). We also show here that
other possible interpretations of the X-ray emission (unlikely a priori)
do not match our data. 
After a short discussion
(section 5), section 6 summarises the results. 

We adopt $\Omega_\Lambda=0.7$, $\Omega_m=0.3$ and $H_0=70$ km s$^{-1}$
Mpc$^{-1}$.  The scale, at $z=1.9$, is 8.4 kpc arcsec$^{-1}$.
Magnitudes are quoted in the photometric system in which they
were published (Vega for near-infrared photometry, AB for optical 
photometry), unless stated otherwise.

\begin{figure*}
\centerline{\psfig{figure=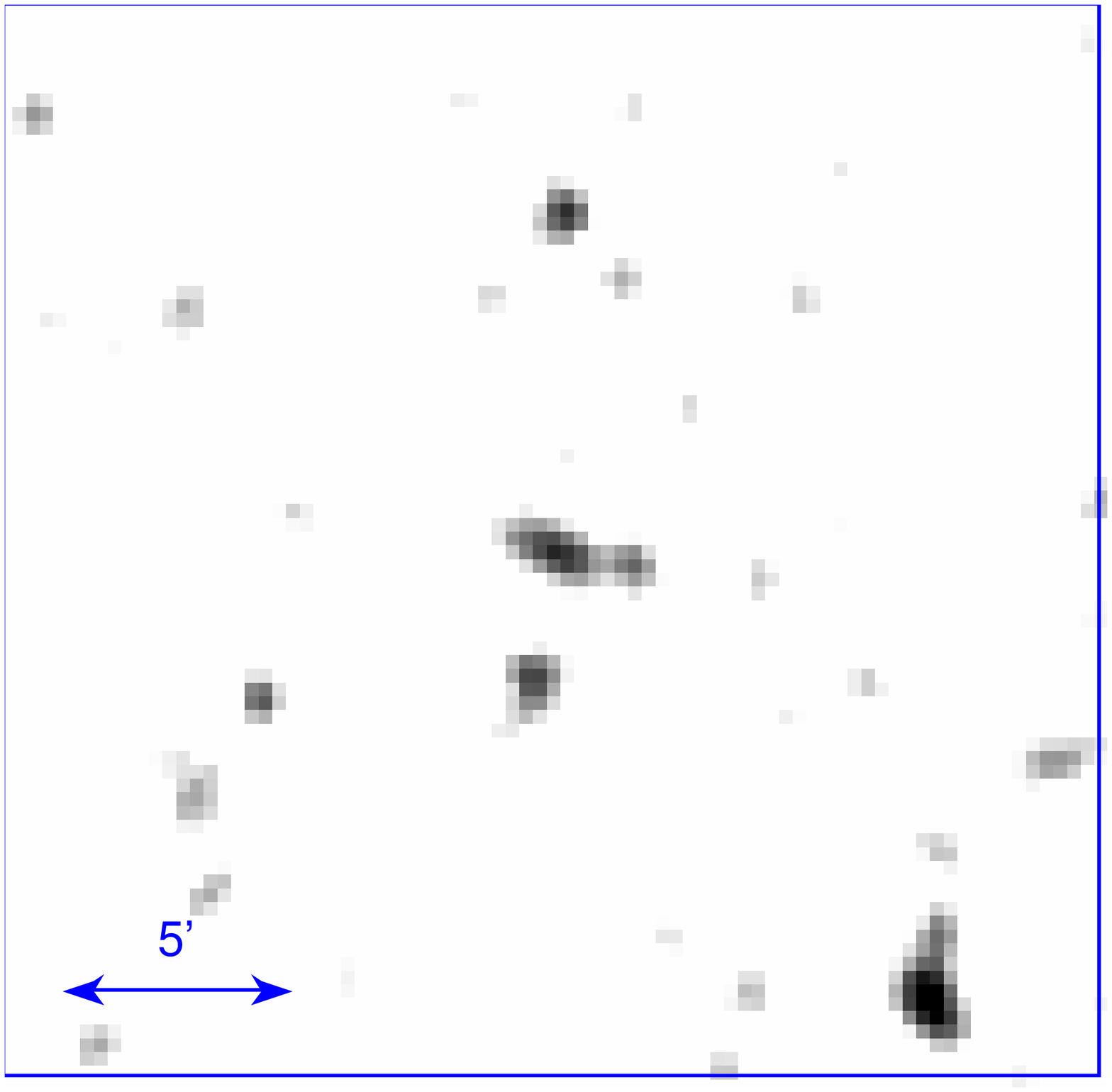,width=5truecm,clip=} \quad %
\psfig{figure=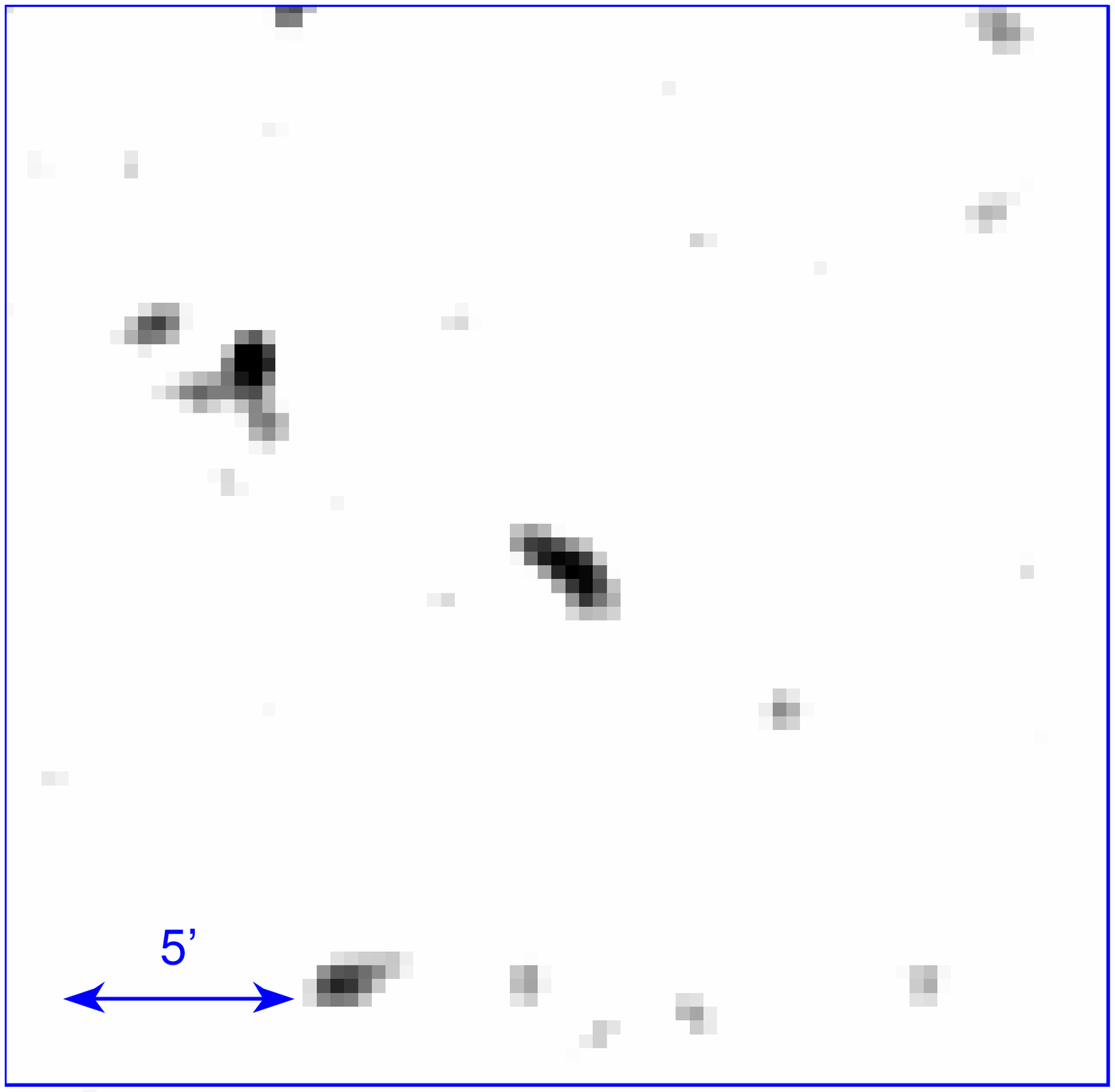,width=5truecm,clip=} \quad%
\psfig{figure=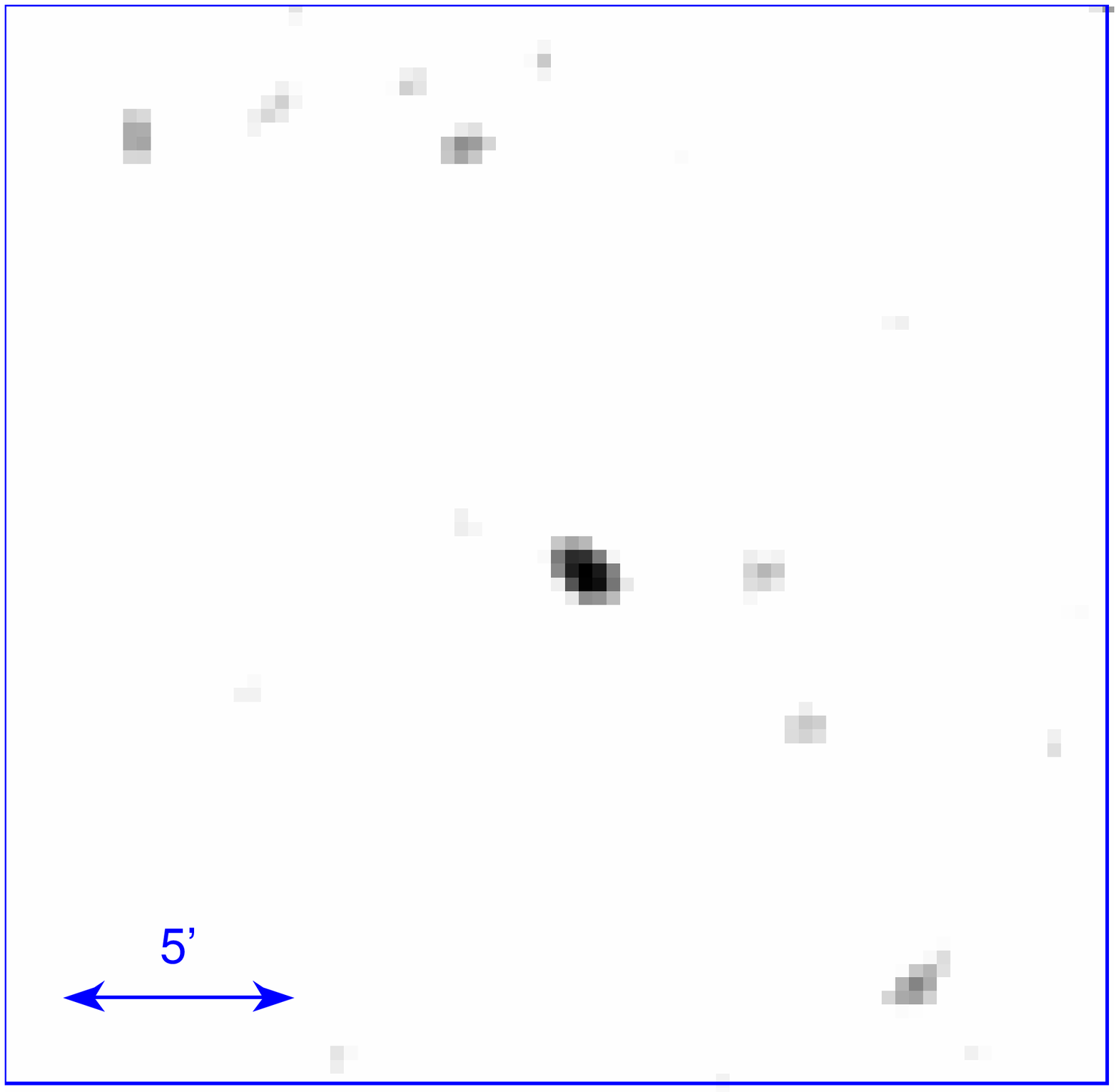,width=5truecm,clip=}}
\caption[h]{Number density image of a region of
146 Mpc$^2$ (0.16 deg$^2$) area centred on JKCS\,041. 
In the left panel, only $2.1<J-K<2.5$ mag galaxies are considered, and
all the ancillary photometry ignored. This shows the original cluster
detection. In the central panel, we have discarded foreground galaxies,
as identified by their spectral energy distribution (SED), using spectroscopy, 
optical and near-infrared photometry. In the right panel,
we only keep galaxies with SEDs similar to the
Grasil 1.5 Gyr old ellipticals at $z=1.9$ and we also use Spitzer
photometry.  
JKCS\,041, at the centre of each panel, is clearly detected in all images, 
independently of the filtering applied. The images have been smoothed with
a Gaussian with $\sigma=54$ arcsec for displaying purpose. A 5 arcmin
ruler is also marked. The large amount
of white/light-gray space in each panel qualitatively shows that the detection
is unlikely to result from chance background fluctuations. North is up,
East is to the left.
}
\end{figure*}

\section{Near-Infrared UKIDSS data and cluster discovery}

JKCS\,041 was initially detected in 2006 using $J$ and $K$
UKIRT Infrared Deep Sky Survey (UKIDSS, Lawrence et al. 2007) 
Early Data Release (Dye et
al. 2006) as a clustering of sources of similar colour using our own
version (Andreon 2003; Andreon et al. 2004a,b) of the red-sequence
method (Gladders \& Yee 2000). UKIDSS data used here
are complete (5 sigma, point
sources, 2 arcsec aperture) to $K=20.7$ mag and $J=22.2$ mag
(Warren et al. 2008). 

Basically, all non-Bayesian cluster detection algorithms (including
ours) compute  a $p-$value (called significance in CIAO, the Chandra
software package, and  detection likelihood in  XMMSAS). This is
sometimes described as the  probability of rejecting the null
hypothesis ``no cluster is there", for a set of parameters.  In our
case, these parameters are sky location, red-sequence colour, cluster
size, red-sequence colour width and limiting magnitude. The
N-dimensional volume of the parameter space is explored iterating on
all (or many) values of the parameters. For the following analysis, we
used a  regular grid of parameter values.

JKCS\,041 is detected with a $p-$value of $10^{-11}$ at $J-K=2.3$ mag
and ra,dec=(36.695,-4.68) on a scale of 1 arcmin and with a
red-sequence width of 0.2 mag.  This would correspond to $\sim 6.5
\sigma$  detection in the classical hypothesis testing sense (assuming
a Gaussian distribution). The left panel of Fig. 1 shows the spatial
distribution of the number density of galaxies with $2.1<J-K<2.5$
centred on JKCS\,041. 
The large amount
of white/light-gray space qualitatively shows that the detection
is unlikely to be produced by chance background fluctuations.

Since the discovery of JKCS\,041, the UKIDSS $4^{th}$ data release
catalogue (Warren et al. 2008) has been released. We make use of that
data here, together with images that we mosaiced from Early Data
Release stacks (Dye et al. 2006).

The detection just described makes a very limited use of the available
data. In the next sections we show how much
we can infer from optical, near-infrared and
IRAC data to characterise JKCS\,041 in absence of the
X-ray data. These methods are of interest for clusters for which X-ray 
data are not available, or, even worse,
their expected X-ray flux is too low to be detected in a
reasonable exposure time. The analysis of JKCS\,041 X-ray
observations is presented in Sec. 4.

\section{What can we learn about JKCS\,041 without X-ray data?}

Multiwavelength coverage is available for
JKCS\,041 from a) the Canada-France-Hawaii Telescope
Legacy Survey (CFHTLS, hereafter) Deep Survey 1 field, b) the Swire (Lonsdale 
et al. 2003) fields, and c) the VVDS $2^h$ spectroscopic field.

CFHTLS gives images (available at the Canadian Astronomy Data Centre) and 
Ilbert et al. (2006) give catalogues in five bands: $u^*,g',r',i',z'$.
Spitzer gives images and catalogues in four bands ([3.6], [4.5],
[5.8] and [8]). We used Spitzer images, as presented in Andreon (2006a),
and the Spitzer catalogue, as distributed by Surace et al. (2005), the latter
taken (in place of Andreon 2006a catalogue) to make our work easier to reproduce. 
Le F{\`e}vre et al. (2005) give VVDS spectroscopic redshifts in the general
area of JKCS\,041 for about $10^4$ galaxies. However, cluster members
are not included in the spectroscopic catalogue, because they are
fainter than the VVDS limiting magnitude ($I_{AB}=24$). We make use of only those
VVDS redshifts that are considered reliable ($flag\ge2$).

In order to enhance the cluster
detection, we used the above data in two ways: a) 
to remove foreground objects from the sample, i.e. 
galaxies that are at much lower (photometric) redshift or, b), 
to retain galaxies whose observed spectral energy distribution
(SED, hereafter) fits that of old (red) galaxies at
$z=1.9$.

\subsection{Removing foreground objects}

By combining near-infrared photometry, optical photometry and VVDS
spectroscopic data we can remove foreground galaxies. We
flagged as foreground every galaxy whose SED
includes a detection in at least five filters and 
matches the SED of at least two galaxies with VVDS redshift $z<1.6$. We require a match to two
VVDS galaxies (instead of just one) to make our flagging more robust against VVDS galaxies
with (potential) bad photometry 
in our catalogues (for example, because of a deblending  problem). We define two SEDs
as matching if they differ by less than 0.05 magnitues in each colour index
($u^*-g',g'-r',r'-i',i'-z',z'-J,J-K$). The advantage of this approach, compared to
photometric redshift estimates, is that all systematics (due to seeing effects,
photometric calibration, templates mismatches, etc.) cancel out in the
comparison. The trial sample is from the very same image (and catalogue) as the
measured sample, and shares its idiosyncrasies. Furthermore, the above
approach does not suffer from redshift degeneracies, which affect photometric
redshifts ($z_{\rm phot}$), which attempt an inversion by deriving $z_{\rm phot}$ from
SEDs.

The central panel of Fig. 1 shows the spatial distribution of the galaxies whose
SEDs do not resemble VVDS $z<1.6$ galaxies. The noise in the map
is reduced after the removal of these foreground galaxies, while JKCS\,041 is still prominent. 
Since galaxy
positions are not used to decide whether a galaxy is in the foreground, 
the spatial structure we see in the image is not a spurious feature of the 
foreground removal.

\begin{figure}
\psfig{figure=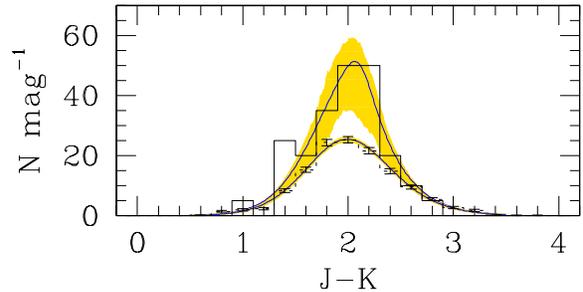,width=8truecm,clip=}
\caption[h]{
The $J-K$ colour distribution of galaxies within
1.5 arcmin (750 kpc) of the cluster centre is shown by the solid histogram. 
The same distribution measured in the control field (a $1.5<r<15$ arcmin annulus)
and normalised to the cluster area is shown by the dashed histogram with
error bars. Foregound galaxies were removed in both cases.
A clear excess is seen at $J-K\sim1.9-2.2$ mag. Error bars mark approximate
errors, whereas 68 \% highest posterior credible bounds are marked
by the shading. The models represented by the solid curves are described 
in the text.}
\end{figure}

Figure 2 shows the colour distributions of galaxies not resembling VVDS
galaxies at $z<1.6$, and brighter than $K=20.7$, for two different regions: 
within a 1.5 arcmin radius 
(750 kpc) from
the cluster centre, and in a control annulus from 1.5 to 15 arcmin in radius. The latter
distribution was normalised to the area of the cluster region. The
 control histogram gives the expected number of galaxies unrelated to the cluster
(those in the foreground but not identified as resembling $z<1.6$ VVDS galaxies, plus those in the background). More
precisely, this second histogram is the maximum likelihood estimate of 
the true value of the background. To determine the significance of the cluster detection, we used the Bayesian methods introduced in Andreon et al. (2006b) and used in Andreon et al.
(2008) to model the same problem. This removes the approximation of the maximum
likelihood estimate (i.e. we allow the background to be as uncertain as the data allow, instead of assuming a perfect knowledge of it). The colour distribution of the
background data was fit with a Pearson type IV distribution (that allows a larger
flexibility than a Gaussian, allowing non-zero skewness and excess kurtosis)  
and the colour distribution of galaxies in the cluster region were fit with
the background model plus a Gaussian. The intensities of these processes are
Poisson, with the obvious constraint that the background rate (i.e. intensity
divided by the solid angle) in the cluster region and the control region are the same. As priors, we took uniform distributions for all
parameters in their plausible ranges and zero outside them (avoiding, for example,
normal distributions with negative variances, a colour scatter
lower than the colour errors, negative numbers, etc.). A
Marchov-Chains Monte Carlo (Metropolis et al. 1953)  with a Metropolis et al.
(1953) sampler was used to perform the stochastic computation. 

\begin{figure}
\psfig{figure=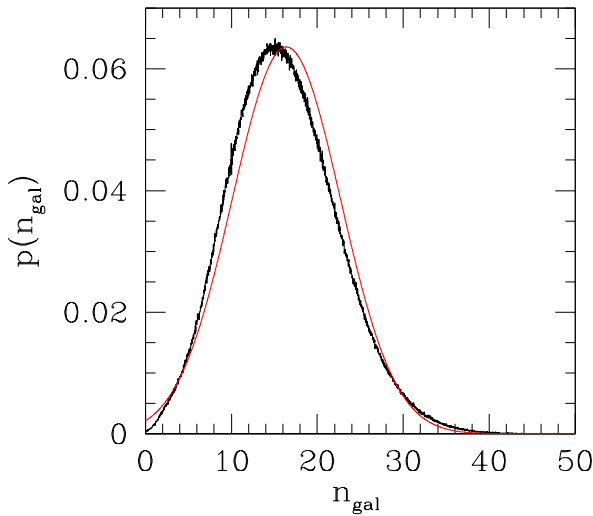,width=8truecm,clip=}
\caption[h]{
The black jagged curve shows the posterior probability of the number of JKCS\,041 members.
Also plotted is a Gaussian with the same mean and standard deviation (red curve). Note that the Gaussian is a poor approximation to the probability distribution at low $n_{gal}$ values, demonstrating that a Gaussian approximation cannot be
used to establish the detection significance in terms of a number of sigma's.
The posterior value at $n_{gal}=0$, normalised by the prior value at that value (about 0.02), gives the relative probability of the two hypotheses:
``no cluster" vs "cluster". As it is obvious even by inspection of the figure, 
this ratio is near to zero.
}
\end{figure}

The (blue) curves and yellow shaded regions in Fig. 2 show the best-fitting models and
the highest posterior 68 \% confidence intervals, respectively. There is a clear excess
above the background, at $J-K\sim 2.1$ magnitudes. This is quantified in Fig. 3, which
shows the probability distribution of the number of cluster galaxies; there are about
$n_{gal}=16.4\pm6.3$ cluster galaxies. We note that the probability distribution is not
normal (the red curve is a Gaussian with matching mean and standard deviation),
especially at low $n_{gal}$ values.  The peak of the colour distribution, $J-K\sim 2.1$
mag, is slighly bluer than the detection colour, $J-K= 2.3$ mag, because of the presence
of blue members, and because we attribute, for this plot only, $J=22.2$ mag to
galaxies undetected in the $J$ band.

A second peak in the colour distribution is visible
at $J-K\sim1.4$ mag. Our model does not account for 
this feature (supposing it to be real, and not to be another cluster/group
on the line of sight), because we have not allowed two peaks in
the colour distribution of cluster galaxies. If we
added a second Gaussian component in the cluster model, the total number of
cluster galaxies is $19.3\pm6.3$, i.e. there are three additional galaxies,
not accounted for in our simpler cluster model. With this exception (a half
sigma difference), all of the 
inferences based on the simpler model are unchanged. 

\subsection{Detection Probability}

In Sec. 2 we computed the cluster ``detection significance" and we found
$10^{-11}$. This number is technically known as $p-$value,  and it is
the quantity most often quoted in astronomical papers to measure the
strength of a detection. By definition, a $p-$value is the probability
of observing, under the null hypothesis, a value at least as extreme as
the one that was actually observed.  Readers willing to evaluate the
strength of our cluster detection compared to other cluster detections
should use our $p-$value. The detection of JKCS\,041 cluster is as
``sure" as other published $10^{-11}$ detections.

Readers desiring to use Bayesian evidence are asked to follow us along a longer
path. We have some data (those mentioned above) and we want to know
the relative probability (often called evidence hereafter) 
of two models: one that claims ``a cluster is there",
and another that claims ``no cluster is there". To evaluate this, we simply need to compute
these two probabilities and calculate their ratio.  The two models (hypotheses) are
nested: ``no cluster is there" is a (mathematical) special case of ``at least
one cluster is there", when all clusters have precisely zero members
each\footnote{The likelihood ratio theorem, or statistical tests build on it,
such as the F-test, cannot be used in our case because the tested model is on
the boundary of the parameter space, see e.g. Protassov et al. 2001 or Andreon
2009 for an astronomical  introduction.}. We assume a priori equal
probabilities for the two hypotheses, to express our indifference between the
two hypotheses in the absence of any data. When hypotheses are nested, as in our
case, the ratio above simplifies to the Savage-Dickey density ratio (see
Trotta 2007 for an astronomical introduction). The latter ratio is
computationally easier to calculate, because it is the value of the posterior
at the null hypothesis (i.e. at $n_{gal}=0$ in Figure 3) divided by the prior
probability of that value.  For a uniform prior on $n_{gal}$ between 0 and
50, we find a ratio of $1.9 \times 10^{-2}$. Therefore the probability that a cluster is there
is about 50 times larger than the probability that none is there. This constitutes strong
evidence on the Jeffreys (1961) scale (see Liddle 2004 for an astronomical
introduction to the scale). Our evidence ratio implies that only one in 50
clusters detected at the claimed significance of JKCS\,041 is a statistical
fluctuation. The precise value of the evidence ratio depends only slightly 
on the assumed prior, provided a reasonable one is adopted. Let's assume,
for example an exponential declining prior, $p(n_{gal})\propto e^{-n_{gal}/\tau}$, 
to approximate the fact that nature usually
produces a lot of small objects (e.g. groups) per each large object (e.g. a rich
cluster). With this prior, we found odds of 1 in 50 for all $\tau\ga 5$
 (N.B. smaller values of $\tau$ were not considered as they give very
low probabilities for $n_{gal}>30$, which is clearly inconsistent with 
observations of clusters, e.g. Abell 1958 and Abell, Corwin \& Olowin 1989).

This evidence ratio cannot be compared to $p-$values. 
They are fundamentally different quantities, and their numerical values differ 
by orders of magnitude (nine, in the case of JKCS\,041). Evidence ratios
are not commonly used in the astronomical literature to quantify
the quality of a cluster detection, but we can 
compute them ourselves using published data for the 
REFLEX cluster survey (Bohringer et al. 2004). For this, we will use the property 
mentioned previously that 
the evidence ratio is the ratio of ``not-confirmed"
to ``confirmed" clusters. 
 Clusters in REFLEX 
were selected to be X-ray sources, to positionally match a
galaxy overdensity, and to have a spectroscopic redshift.
The XMM follow-up of a REFLEX  
sub-sample of clusters (REXCESS, Bohringer
et al. 2007) choose 34 clusters in REFLEX
not amongst the worst, and the chosen objects
are said to be representative of REFLEX clusters by the authors.
They found one AGN among 34 objects previously classified
as clusters in REFLEX. 
The implied quality of a REFLEX cluster detection is
therefore 1 in 33, marginally lower than the 1 in 50 computed for the JKCS\,041 
detection. We emphasise that the good REFLEX performance
used X-ray, optical and spectroscopic information on the cluster.
Instead, the ``1 in 50" evidence ratio provided here to measure the
strength of the JKCS\,041 detection does not make any use of the
(available) X-ray data.  The JKCS\,041 detection
is slightly more secure than REFLEX clusters. The introduction of the 
X-ray evidence only strengthens the cluster detection significance, as
JKCS\,041 is 
detected as an extended X-ray source (Sec. 4).
This provides a solid
confirmation of our probability calculus to asses the significance of its detection.

\begin{figure}
\centerline{\psfig{figure=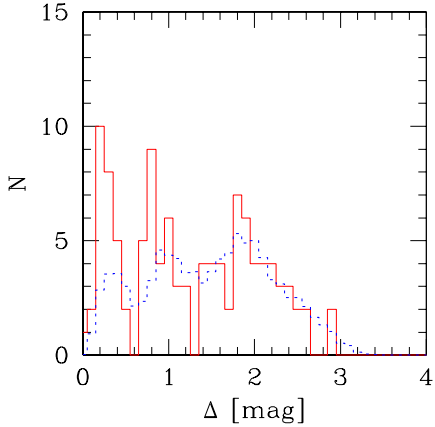,width=6truecm,clip=}}
\caption[h]{
Distribution in the generalised colour $\Delta$ for
galaxies within
1.5 arcmin from the cluster centre (solid histogram), and of the
control field (measured in a $1.5<r<15$ arcmin annulus, dashed histogram), 
normalised to the cluster area. 
A clear excess is seen at $\Delta<1$ mag.}
\end{figure}

\subsection{SED detection using eleven bands}

The second approach to detecting JKCS\,041  uses all eleven photometric bands to
identify  galaxies with a spectral energy distribution similar to a Grasil 
(Silva et al. 1998) $1.5$
Gyr old elliptical galaxy at $z=1.9$ (i.e. $z_f\sim 3.5$ for the adopted
cosmology).

Fig. 4 is similar to Fig. 2: it shows the colour distribution, but for a 
generalised colour, $\Delta$, given by the average distance of the photometry data points 
from the model SED (a $1.5$
Gyr old Grasil elliptical at $z=1.9$).
Had we used just the $J$ and $K$ photometric bands, the x-axis would be $J-K$
(plus an obvious warping, to bring all measurements to a common 
numerical scale) and Fig. 4 would be identical to Fig. 2, except for
the different sample selection.  We consider here
galaxies with photometric data measured to better than
10 \% accuracy in at least 4 bands.
In the cluster
direction there is a clear excess of galaxies having SEDs  similar to the $1.5$
Gyr old Grasil elliptical (i.e. with small values of $\Delta$). For
example, we observe 23 galaxies with $\Delta<0.4$ when 9.09 are expected. The
probability of observing a larger value by chance alone is $3 \ 10^{-5}$.
At first
sight, the observed $\Delta$ values are large, of the order of 0.2-0.3 mag for galaxies in
the leftmost peak. This is expected, however, primarily because  we are sampling the ultraviolet ($1200-3100$
\AA ) with many data points. This means that small differences between the true history
of star formation and the model result in large differences between
observed and model SEDs in this wavelength range.

This SED based approach 
is analogous to the photo-z approach used by
Stanford et al. (2005) to detect the $z=1.41$ ISCS
J1438+3414 cluster. Therefore, had we decided to use SED model fitting as our initial detection method, JKCS\,041 would still have been detected.

The spatial
distribution of these SED-selected galaxies with $\Delta<0.4$ mag is shown 
in the rightmost panel
of Fig. 1. JKCS\,041 shows the largest numerical overdensity in the survey area 
($53 \times 53$ arcmin$^{2}$).  

\begin{figure}
\centerline{\psfig{figure=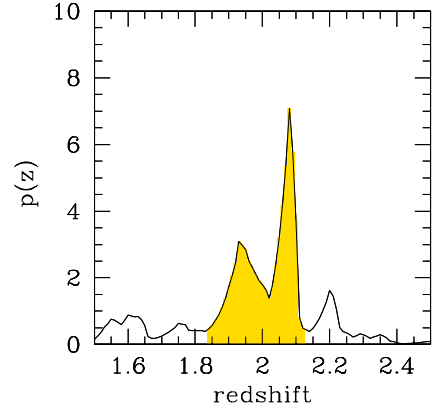,width=6truecm,clip=}}
\caption[h]{Photo-z redshift probability distribution, based on
11 bands photometry. The shortest 68 \% credible interval
is marked with a (yellow) shading.}
\end{figure}

\subsection{Cluster redshift}

Due to the
faintness of JKCS\,041 galaxies ($I\sim 25$) and their weak spectral
features in the (observer-frame) optical, 
we failed to measure spectroscopic redshifts 
even after an exposure of 12 hours on FORS2 at VLT (progr. P277.A-5028). 
We therefore address
the determination of the cluster redshift by using galaxy colours.

The original cluster detection colour (sec. 2), $2.1<J-K<2.5$ and
the peak of the colour distribution (sec. 3.1), both
imply $z_{phot}\approx 1.9$ either assuming a model spectral energy 
distribution (e.g. Bruzual \& Charlot 2003, or those detailed in the
UKIDSS calibration paper, Hewett et al. 2006) or 
by comparison with $J-K$ colour of
red galaxies at $z_{spect}\approx 1.9$ (e.g. Kriek et al. 2008).

We now use our SED approach of Sec. 3.3 to compute the photometric cluster
redshift and its uncertainty. For each redshift value we compute
the expected Grasil SED for $1.5$
Gyr old elliptical galaxy and we compute how many galaxies
match ($\Delta<0.4$) this SED in the cluster direction
(a circle of 1.5 arcmin), $n_{on+off}$, and
in a reference line of sight, $n_{off}$, 
the latter estimated in a corona centred on the cluster and
with inner and outer radii of 1.5 and 15 arcmin (and scaled
by the solid angle ratio). Assuming that these 
count processes are Poisson distributed, the posterior probability
$p(z|n_{on+off},n_{off})$ is computed assuming uniform priors,
following simple algrebra
(e.g. Prosper 1989, Kraft 1991, Andreon et al. 2006). The posterior
distribution is plotted in Fig. 5, for $z>1.5$ because $z<1.6$
is already ruled out by having detected the cluster after filtering
out $z_{phot}<1.6$ galaxies (sec. 3.1).
The posterior probability distribution has
mean equal to $z=1.98$ and
two peaks, one at $z=1.93$ and one at $z=2.08$. The
68 \% shortest confidence interval, shaded in the figure, is
$[1.84,2.12]$. 

The length of the 68 \% uncertainty interval on redshift, $0.28$,  can also
be estimated from
 the error on the mean
$J-K$ colour of the red galaxies that compose the excess in Fig. 2,
after restricting the sample to $J<22.2$ mag in order
to avoid upper limits. The result is 0.22 mag (the value is derived from
the modelling described in Sec. 3.1). Since $d(J-K)/dz = 1.8$
for old galaxies at $z\approx 2$ (e.g. using 
models listed in Hewett et al. 2006), this gives 
$\sigma_z=0.22/1.8=0.12$ (vs $0.14=0.28/2$ for half 
the 68 \% shortest confidence 
interval).

We emphasise that this is the statistical error. However,
we expect a systematic error due to our use of a model SED,
instead of the true $z\sim2$ old galaxies SED, directly measured 
on the same data used by us (i.e. UKIDSS+CFHTLS+IRAC).
For example, Hewett et al. (2006), describing
the photometric calibration of the UKIDSS survey (from which
we took $J-K$), give two predictions of the colour of an 
elliptical in the UKIRST
photometric system at $z\sim 2$ which differ by 0.1 mag 
($\Delta z = 0.05$).
A similar comparison (Kriek et al. 2008), 
based however on real spectroscopic and $J-K$ measurements, 
displays $\Delta z \sim0.08$ systematic errors. 
At much lower redshift
EDiSCs clusters have photo-z systematics of $\Delta z \sim0.1$ 
(White et al.  2005), similar to RCS clusters prior to
photometric redshift recalibration (Gilbank et al. 2007). 

 This source of error has little effect on the 
width of the posterior redshift distribution: 
if we model this source of uncertainty with a top-hat
filter of  width $\Delta z=0.2$, (i.e. $\pm 0.1$), 
the 68 \% confidence interval of JKCS\,041 redshift
is $[1.86,2.18]$, almost identical to what intially derived
($[1.84,2.12]$).

The redshift uncertainty has little effect on our results. It
does not change at all the probability that we detected
a cluster. Our filtering technique (sect. 3.1) excludes a 
redshift similar to that of the
``now" second most distant cluster of $z=1.45$ (Stanford et al. 2006).       
It affects the value of cluster core radius (sec. 4.1) by less
than 1 per cent, because the angular distance is almost constant with
redshift at $z\approx2$.
It introduces an uncertainty on the cluster X-ray luminosity (sec. 4.2) by 
about about 15 per cent (if $\delta z \sim 0.1$), only three times 
the uncertainty implied by the uncertainty on $\Omega_m$ alone, and
by a negligible quantity for the study of $L_X-T$ scale relation, because
the mentioned 15 per cent error is
about five time smaller than the $L_X$ scatter at a given $T$
(e.g. Stanek et al. 2006). 
The redshift uncertainty represents a minor source of error on the cluster
temperature (sec. 4.2.).

When needed for intrinsic quantities,
we adopt the lower-redshift peak of the distribution,
$z=1.9$ as the cluster redshift, in place of the
posterior average value, $z=1.98$, which gives   
conservative estimates of the cluster redshift, X-ray
luminosity, and cluster mass.

\begin{figure}
\centerline{\psfig{figure=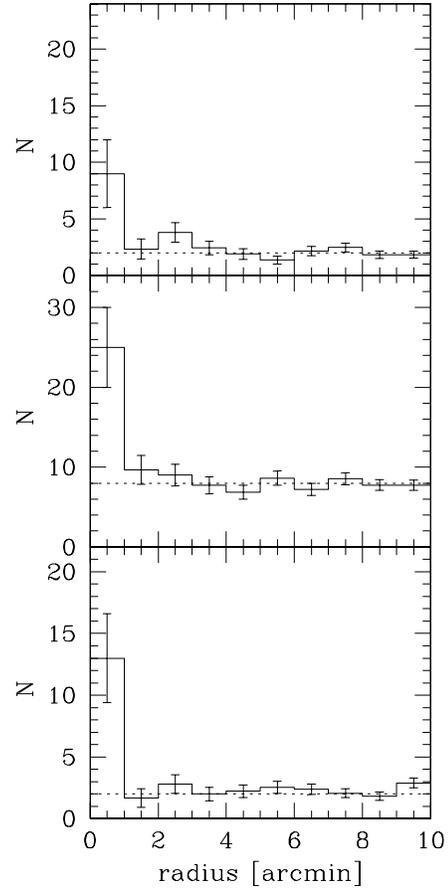,width=6truecm,clip=}}
\caption[h]{Number density radial profile of galaxies selected according
to $J-K$ colour (top panel), dissimilarity from VVDS galaxies SED 
(central panel), similarity to old galaxies at $z=1.9$ (bottom panel).
In all three cases, more galaxies in the cluster line of sight are observed
than in adjacent directions.}
\end{figure}

\subsection{A robustness check}

We now revisit our cluster detection, using a slighly more stringent  
magnitude limit and a simplified analysis than applied in 
previous Sections. 

We first consider galaxies with $K<20.1$ (well within the K-band limit).
Those with $J-K\le2.1$ are brighter than the completeness limit in the $J$ band,
$J=22.2$, and removed because they are not of interest.  All other 
$K<20.1$ galaxies, regardeless of their actual detection in the $J$ band,
will have $2.1<J-K<2.5$ (we expect minimal 
contamination from $J-K>2.5$
sources).  Their number density radial profile, shown in the
top panel of Fig. 6, indicates an excess in the 
inner 1 arcmin, where 9 galaxies are found when 2 are expected. This implies a 
significance ($p-$value) of $2 \ 10^{-4}$ (about a
$3.7$ $\sigma$ detection), lower than reported in section 2, but still
generally acceptable.  

We then consider galaxies with $K<20.5$ mag. We remove
from the sample galaxies with SED matching any pair
of $z<1.6$ VVDS galaxies in at least three bands with
good photometric quality
($<0.2$ mag). Again, the radial distribution 
of the remaining population (middle panel of Fig. 6)
shows an excess in the inner $1'$:  we observe 25 galaxies when 8 are expected.
The $p-$value (significance, detection likelihood) 
is $10^{-6}$, a $5 \sigma$ detection.

\begin{figure*}
\psfig{figure=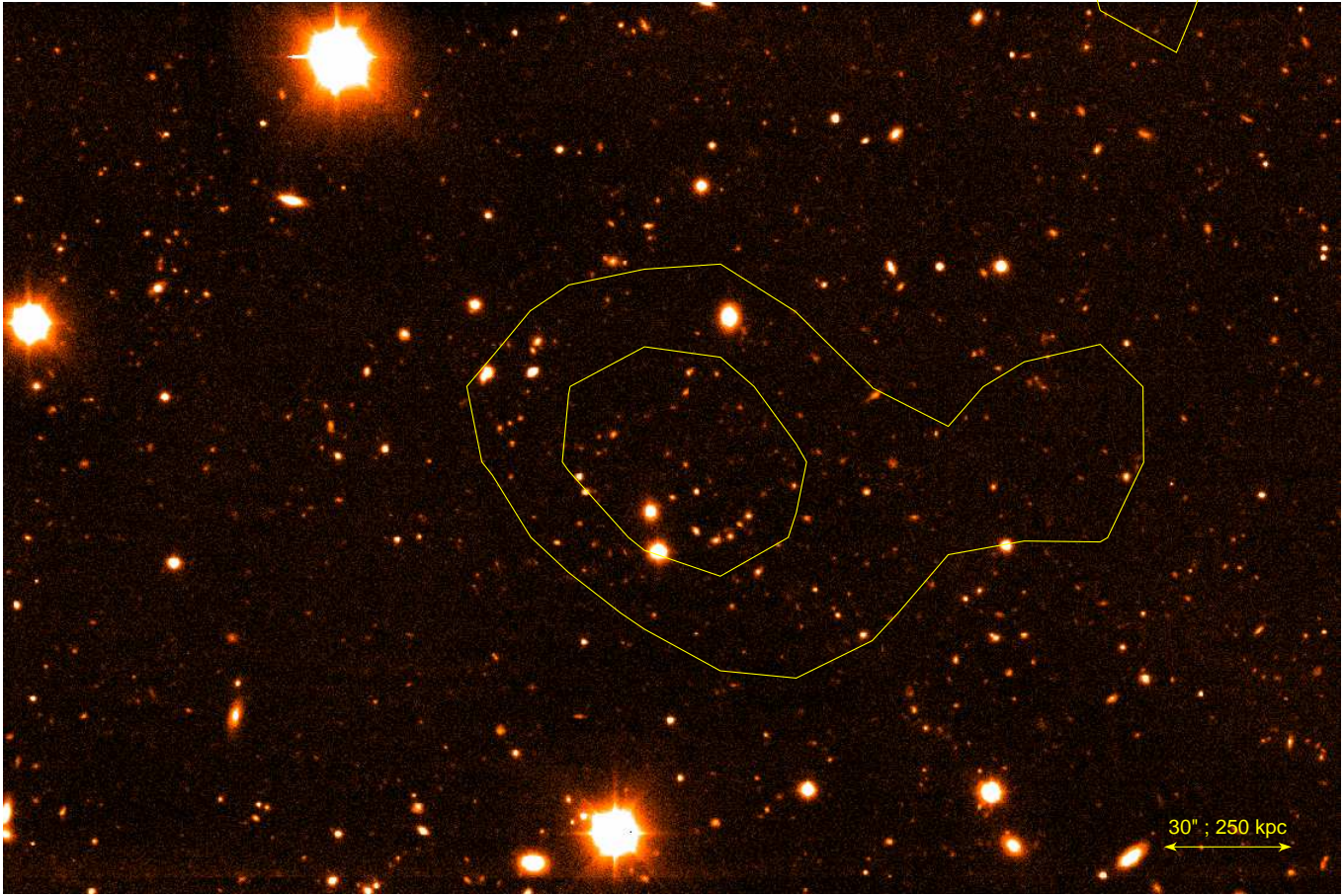,width=18truecm,clip=}
\caption[h]{$z'$ image. 
Contours of equal number density of
galaxies with $2.1<J-K<2.5$ mag (irregular yellow contours).
North is up and East is to the left.}
\end{figure*}

Last, we consider galaxies with $K<20.5$ mag,  good photometric quality
($<0.2$ mag) on at least 6 bands,  whose  SED are similar to a $z=1.9$
Grasil 2.5 Gyr old galaxy (sect. 3.3).  Fig. 6 (bottom panel) shows  13
galaxies in the inner 1$'$ when 2 are expected, giving a 
$p-$value (significance, detection likelihood) of 
$2\ 10^{-7}$, a $5.3 \sigma$ detection.

Although with reduced statistical significance, due to the
reduced sample, we confirm a spatial concentration of 
objects of similar colour or SED within 1$'$
from the cluster centre. 
The typical colour of the excess population
is $J-K\sim 2.3$ mag, either because this is the selecting
colour (first method), because this is the colour of
the bulk of the population after foreground removal
(second method)
or because it is the colour of the template SED 
used to select the galaxies (third method).

Figure 7 shows a large area around JKCS\,041 in the VLT $z'$ band,
where the galaxy overdensity is highlighted by the yellow
contours. A true--colour image ($z'JK$) of a slightly smaller
region around JKCS\,041 is given in Figure 8.

\subsection{A single cluster or a blend of smaller structures?}

In the previous section we showed that the detection of JKCS\,041  
is significant and not just a statistical fluctuation.
However, the detection algorithms used do not provide constraints on the
size of the
detected structure along the line of sight. In particular, they
do not distinguish between the observation of a single 
cluster-size object, or a projection  
of two (or several) groups.  
This limitation is common to several
other cluster detection methods. The line of sight kernel of
cluster detection by
gravitational shear is about 1000 Mpc, basically because the 
detected signal changes little on these scales.
Sunyaev-Zeldovich (SZ) cluster searches have an even larger kernel, $\delta z=1$,
basically because the signal depends on the angular size distance, which flatens off at
$z\sim0.4$. In fact, SZ cluster surveys are likely to be confusion-limited
at masses below $10^{14} \ h^{-1}$ M$_\odot$ (Holder et al. 2007),
below this threshold cluster detections will frequently be blends of
clusters along the line of sight. The same is
partially true at higher masses, especially considering that the confusion
error is highly non-Gaussian with a long tail to the positive side.

\begin{figure}
\centerline{%
\psfig{figure=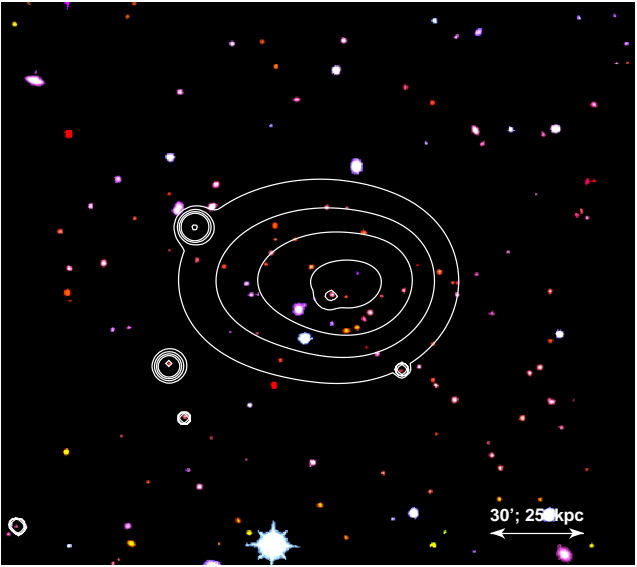,width=8.5truecm,clip=}%
}
\caption[h]{True colour ($z'JK$) image. 
Contours of the adaptively smoothed X-ray emission detected by Chandra 
(white contours).
North is up and East is to the left.}
\end{figure}

Red-sequence-like
cluster searches have a redshift kernel that is given by the  
photometric error of the colour
divided by the derivative with redshift of the model colour.
This implies $\delta z \sim 0.02$ when photometry S/N is high,
the Balmer break is well sampled by the filter pair and filters are taken
near to the break (see
Andreon 2003 and Andreon et al. 2004 for an observational assessment
at $z<0.3$ and $0.3 \la z \la
1.1$, respectively).
$\delta z \sim 0.02$ error at $z=1.9$ (which are very optimistic) would still imply
a resolution of $\approx 30$ Mpc along the line of sight. This is too large
to discriminate, for example, a single cluster of size of $\la 3$ Mpc from 
two structures separated by 10 Mpc. Therefore, red-sequence detected
clusters are also prone to confusion, as we quantify below.

The posterior odds (probability ratio) that a detection is a single object or 
a blend of two objects each carrying about half the total mass
is given by the ratio of the probability of observing one
object, $p(1|\lambda_1)$ over the probability of observing  two objects, 
$p(2|\lambda_2)$, in the given volume, multiplied by the relative a priori
probabilities of the two hypotheses. The latter ratio is taken to be 1 to formalise our
indifference in absence of any data. $p(i|\lambda_i)$, with $i=1,2$
are assumed to be Poisson distributed.
$\lambda_1$ and $\lambda_2$ are 
the average volume density of objects (clusters) at the redshift of
interest, assumed to be a Jenkins et al. (2001) mass function.
The volume is given by 
$1.84\le z \le 2.12$ (the 68 \% confidence interval) and 
$\Delta \alpha = \Delta \delta =60 $ arcsec
(a larger separation on the sky would mean that the two structures 
could be distinguished). We also adopt  
a power spectrum shape parameter $\Gamma=0.6$ and $\sigma_8=0.9$. 
To convert the mass of JKCS\,041
from $M_{500}$ (derived in Sec. 4.2) to the virial mass, we assume a Navarro, Frenk
\& White (1997) profile of concentration 5. We
find $p(1|\lambda_1)/p(2|\lambda_2)= 4.1 \ 10^{3}$. 
Our result
implies that, on average, one blend occurs every 4100  
detections similar to JKCS\,041. For three or more
objects the odd ratio is even larger. 
The odds
do not change appreciably if we allow blends of similar, but not
identical mass (e.g. up to a mass ratio 1:4). 
Thus, there
is ``decisive" evidence in favour of a single object, when
the evidence is measured on the Jeffreys (1961) probability 
scale. 
We emphasise once more that quoted probabilities are not $p-$values,
and that probabilities (quoted here) cannot be compared to  
$p-$values (often quoted in other studies).

Continuing our survey on cluster detection methods, the size of a structure
can obviously also be assessed with  spectroscopic data, because even a few
galaxies at small $\delta z$ (of the order of 1000 km/s) are sufficient to
establish with reasonable confidence that most of a given structure  is of
cluster size. Of course, one
should also account for the non negligible possibility that a number of
concordant redshifts are found by chance (see e.g. Gal et al. 2008). For
example, in the JKCS\,041 direction, we found 9 galaxies with redshift
$z\sim0.96$ within 2 arcmin from the cluster centre. However, the same
number of concordant redshifts
is found in almost every other region of the VVDS area ($\approx 40 \times
40$ arcmin) that had similar sampling rate (the redshift survey is not
spatially uniform). This is just one of the redshift spikes in the general
JKCS\,041 area (Lefevre et al. 2005), and is not the redshift of every
cluster in the area.

\begin{figure}
\psfig{figure=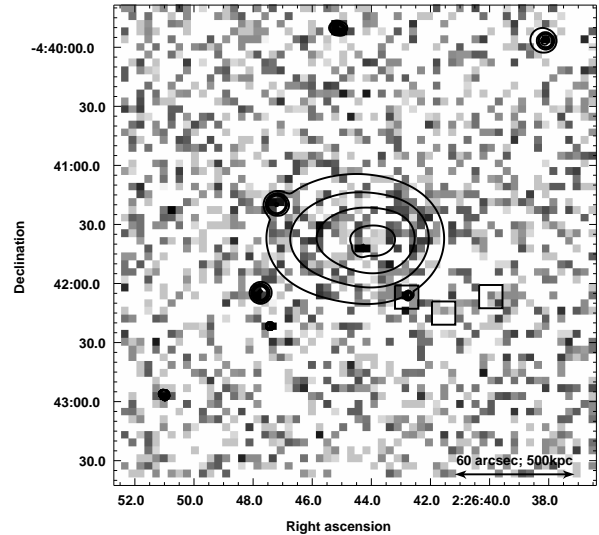,width=8.5truecm,clip=}
\caption[h]{
A background subtracted, exposure corrected [0.3-2] keV Chandra X-ray 
image of JKCS\,041, binned
to 4 arcsec pixels. 
The image is overlaid with contours of the X-ray emission after adaptive
smoothing so that all features are significant to at least the 3 $\sigma$ level. 
The faintest contour was chosen to closely approximate the region enclosing
the pixels for which the smoothing kernel contained a signal above the 3
sigma threshold.
The square boxes
show the positions of 3 radio sources detected in this field, with the box size
indicating the spatial resolution of the radio data (see section 4.3).}
\end{figure}

\begin{figure}
\psfig{figure=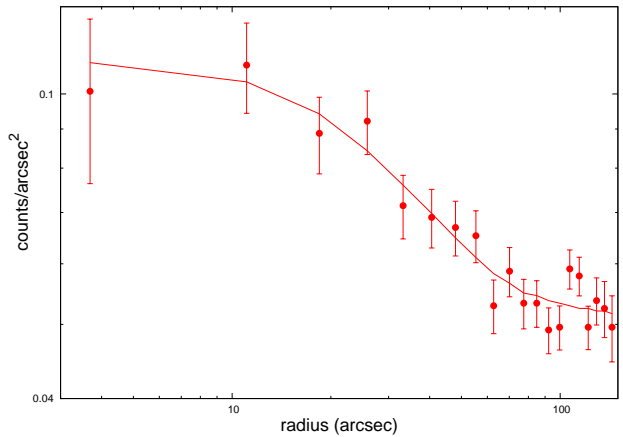,width=8.5truecm,angle=-90}
\caption[h]{Radial profiles of JKCS\,041, and the best fitting 2D model.
}
\end{figure}

\section{Chandra X-ray Observations}

JKCS041 was observed by Chandra for $75$ ks on 2007 November 23 (ObsID 9368),
using the ACIS-S detector. The data were reduced using the standard data
reduction procedures as outlined in Maughan et al (2008). A
preliminary examination of the cluster spectrum showed that the $0.3-2.0$ keV
energy band gave the maximum cluster signal to noise ratio for our image
analysis. An image was produced in this energy band, and is shown in Fig. 9. The
image was then adaptively smoothed so all features were significant to at least
$3\sigma$, using a version of the Ebeling et al (2006) algorithm
modified to include exposure correction. Contours of this smoothed X-ray image
are overlaid on Fig. 9 and on the true-colour image in Fig. 8. The X-ray morphology appears
regular, but this should be interpreted with caution due to the relatively large
smoothing kernel required by the low signal to noise cluster emission. The
$\sigma$ of this Gaussian kernel was $\la20$ arcsec within a radius of $30$
arcsec of the cluster centre. 
Within a 1 arcmin radius from the cluster centre there are $223 \pm 31$ photons in
the 0.3-2 keV band (after subtraction of the background and exclusion of point
sources).

\subsection{X-ray Image Analysis}

The Chandra image of JKCS041 was fit in {\it Sherpa} with a
two-dimensional (2D) model chosen to describe the cluster surface
brightness distribution. The model used was the {\it Sherpa}
implimentation of the 2D $\beta$-profile with an additive constant
background component. The model was constrained to be circular and was
fit to the data using the Cash (1979) statistic, appropriate for low numbers
of counts per bin, including convolution with a model of the Chandra
PSF and multiplication by the appropriate exposure map. Point sources
were masked out of the data and model during the fitting process. The
point source $8$ arcsec to the south east of the cluster X-ray centre
is relatively bright (and already detected in Chiappetti et al. 2005), 
contributing $\sim100$ photons or $\sim30\%$ of
the total cluster and point source X-ray flux in the imaging band (the
other point sources are much fainter). Fortunately, this emission is
easily resolved by Chandra for exclusion from our analysis. 

An advantage of the 2D fitting used here compared with fitting a model to a
one-dimensional (1D) surface brightness profile is that the centre of
such a 1D profile is subject to significant uncertainties in low
signal to noise data like these. Changes in the central position of the 
1D profile have a strong effect on the profile's shape. In a 2D model,
those uncertainties can be explicitly included 
by allowing the centre coordinates of the
2D model to be free parameters. The additional free parameters in our
fit were the core radius $r_c$ of the model, and the source and
background normalisations. The slope $\beta$ of the surface brightness
model could not be constrained by the data and was fixed at
$\beta=2/3$. The best fitting model to JKCS\,041 had central coordinates
of $RA = 02:26:44$ $\pm6$ arcsec and $DEC = -04:41:37$ $\pm4$
arcsec, with a core radius of $36.6^{+8.3}_{-7.6}$ arcsec 
($307^{+70}_{-64}$ kpc). Figure 10
shows a radial profile of the data and the best fitting 2D model. Note
that this is simply for visualisation purposes, the model was not fit in this
space. This figure demonstrates convincingly that the $r_c=37$ arcsec
X-ray emission is extended with respect to the $0.5$ arcsec Chandra PSF.
The cluster core radius, about 300 kpc, is in the range of values observed
for local clusters.

\begin{figure}
\psfig{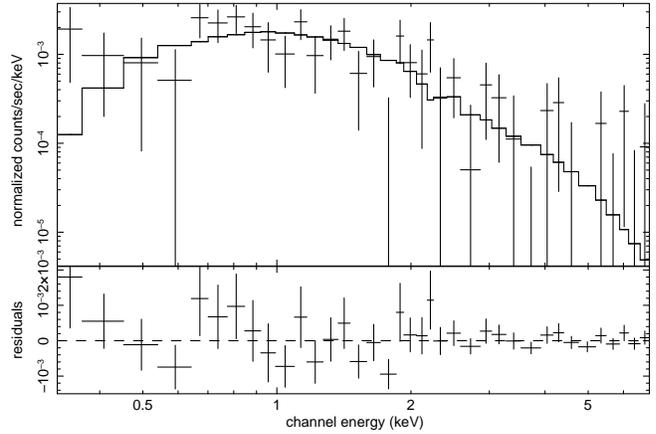}
\caption[h]{The Chandra X-ray spectrum of JKCS\,041 and the best fitting model are shown in the top panel, with the residuals from the model shown in the bottom panel. The spectrum is binned
for displaying purposes.}
\end{figure}

\begin{figure}
\centerline{\psfig{figure=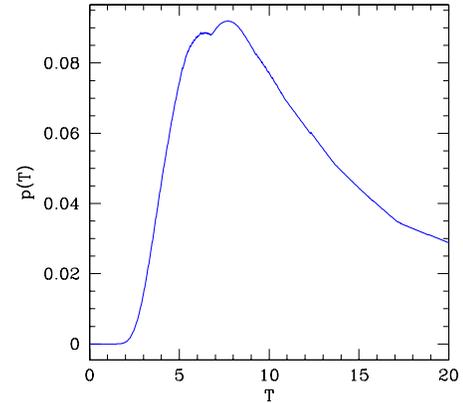,width=6truecm,clip=}}
\caption[h]{Temperature likelihood of JKCS\,041.
}
\end{figure}

\subsection{X-ray Spectral Analysis}

Our spectral analysis procedure was chosen to match that of Pacaud et
al. (2007) to allow direct comparison with their $L_X-T$ relation, presented
in Andreon et al. (in preparation). In
summary, a cluster spectrum was extracted from an aperture of radius
$40$ arcsec  (336 kpc) (with point sources excluded), chosen to maximise the
signal to noise ratio. A background spectrum was extracted from an
annular region around the cluster, sufficiently separated to exclude
any cluster emission (inner and outer radii: 99" and 198";
833 and 1665 kpc). The resulting cluster spectrum contained
approximately 210 source photons in the $0.3-7.0$ keV band used for spectral
fitting, with a signal to noise ratio of $7.4$. The source
spectrum was fit with an absorbed APEC (Smith et al 2001)
plasma model, with the absorbing column fixed at the Galactic value
($2.61\times10^{20}$ cm$^{-2}$), the metal abundance relative to Solar
fixed at $0.3$ and the redshift of the plasma model fixed at
$1.9$. The spectrum was grouped to contain a minimum of $5$ counts per
bin and the model was fit to the background-subtracted data using the XSPEC C-statistic.

The best fitting spectral model had a temperature of
$7.6^{+5.3}_{-3.3}$ keV (plotted in Fig. 11) and gave an unabsorbed
bolometric X-ray flux of $1.93\times10^{-14}$ erg cm$^{-2}$ s$^{-1}$ 
(corrected for
area lost to point sources in the spectral aperture). This temperature
was used to estimate $R_{500}=0.52$ Mpc (the radius within which 
the mean density is 500 times the critical density at
the cluster redshift), 
using the scaling relation of Finoguenov et al (2001) 
as given in Pacaud et al (2007; their equation 2). The best
fitting 2D surface brightness model was then used to scale the
observed flux from the spectral aperture to this radius, including
correction for point sources. The bolometric luminosity within
$R_{500}$ was thus found to be $(7.4\pm1.7)\times10^{44}$ erg s$^{-1}$, with
these errors including the uncertainties on the normalisation of the
spectral model, on its temperature and on the core radius of the
2D model used for the aperture correction. This is consistent again
with Pacaud et al (2007) with the exception that we do not include the
uncertainties on the surface brightness model slope, which is
unconstrained by our data. 

Finally, the temperature of JKCS\,041 can be used to estimate  the cluster's
mass. For consistency, we can simply use the  definition of $R_{500}$ given above
to yield $M_{500}=2.9^{+3.8}_{-2.4}\times10^{14}M_\odot$, under the (strong)
assumption that the Finoguenov et al. (2001) relation holds at $z=1.9$. 

\subsection{Testing a non-thermal origin for the X-ray emission}

Fabian et al. (2001, 2003) interpreted the extended X-ray source coincident with
the powerful radio source 3C 294 at $z = 1.768$ as due to CMB photons scattered
at high energy through inverse Compton scattering on non-thermal electrons
produced  by this radio source. This is the only known case of such a
phenomenum.  
In the case of JKCS\,041, the X-ray spectrum alone is unable to rule out a non-thermal emission model, but
the latter is an unlikely interpretation for several reasons. 
Firstly, there is no significant radio source present in the cluster core.  
Bondi et al (2003; 2007) present VLA and GMRT radio
observations of this field as part of the VVDS-VLA field. The three sources
detected close to JKCS\,041 are marked on Fig. 9. None of the radio sources appear
associated with the extended X-ray emission, although the eastern-most source is
associated with a faint X-ray point source with an optical counterpart

Furthermore, the morhopology of the X-ray emission in 3C 294 was found
to be hourglass shaped, and contained within a radius of 100
kpc. Within the limits of the Chandra data for JKCS041, there is no
evidence for irregular X-ray morphology, and the X-ray emission is
detected to a radius of $\sim500$ kpc (see Fig. 10).

\subsection{Is JKCS\,041 a filament?}

We now consider the possibility that JKCS 041 is a filament viewed
along the line of sight. In short, the X-ray data rule out this
possibility on two fronts: the gas is too dense and too hot.

The density implied by the X-ray emission (as determined from the normalisation of the best fitting spectral model) if the
emission were due to a cylindrical filament of uniform density gas of
length $10$ Mpc along the line of sight and radius $0.326$ Mpc (our
spectral aperture) is
$n_e=4.4\pm0.4\times10^{-4}$ cm$^{-3}$ (giving a mass density of
$1.0\pm0.1\times10^{-27}$ g cm$^{-3}$). This is significantly higher than
the gas densities found in large scale filaments in e.g. the
simulations of Dolag et al. (2006), who found densities typically
much lower than $10^{-28}$ g cm$^{-3}$.

Second, the measured temperature of the gas implies a
deep gravitational potential well, and is inconsistent with the $<1$
keV temperatures predicted in such filamentary structures (e.g. Pierre
et al 2000; Dolag et al. 2006). 
To test how strongly the data can rule out $<1$ keV gas, we need
to calculate the ratio of the probability that $T<1$, as expected
for filaments, and the probability that $T>1$, as usual for clusters. 
The temperature
likelihood (the output of STEPPAR in Xspec), used for the computation, 
is depicted in Fig. 12. If we adopt an uniform prior in $T$, to express
our indifference on $T$ prior to observing any data, over a range abundantly
encompassing all reasonable temperatures (say, from 0.03 to 20 keV),
we found odds of 1 in $\gg \ 10^4$. This demonstrates that the filament hypothesis is 
rejected by the observed X-ray spectrum. 
Fig. 12 shows that this very small probability ratio
is insensitive the precise value (e.g. $2$ keV vs $1$ keV) of the 
temperature threshold adopted to define the two hypotheses.

\section{Discussion}

\subsection{The redshift of the X-ray emission}

The redshift of the X-ray emitting gas,
unless it is directly derived from the X-ray spectrum,
is commonly assumed to be the same of the galaxy
overdensity spatially coincident with it. This is the tacit
assumption made in all but a couple of studies related to
cluster gas intrinsic properties, like X-ray luminosity or temperature.
We make the same assumption and we further check
that no other cluster is in the JKCS\,041 line of sight.

JKCS\,041 lies in a region
rich of photometric data, which have been exploited by several groups
independently.  Using  CFHTLS data, 
Olsen et al. (2007), Grove et al. (2009) and Mazure et al. (2007) 
explored the range up
to $z \sim 1.1$ using several filters and detection
methods. None of their candidate clusters spatially matches
JKCS\,041, the nearest being 6.5 arcmin away. We also found
no matching detection by using $R$ and $z'$ bands and 
the very same algorithm used with $J$ and $K$.
Therefore, we found no evidence of another cluster on
the JKCS\,041 line of sight.

\subsection{JKCS\,041 and cosmological parameters}

The detection of JCSK\,041 is in line with the expectation of a $\Lambda$CDM
model, with $\sigma_8=0.9$ and power spectral shape parameter $\Gamma=0.6$
and a Jenkins et al. (2001) mass function. In the surveyed area
(about $53 \times 53$ arcmin$^2$) and within $1.8<z<2.0$ 
about 2.4 clusters with $M>10^{13.75} \ h^{-1}$ are expected.

In the past, cosmological parameters have been constrained by the
observation of a single high redshift cluster.  The argument used is
quite simple: cosmological parameters that predict in the studied volume
fewer than one cluster more massive than the observed cluster are
disregarded in favour of those that make predictions close to the
observed number of clusters (i.e. one).
Because the first discovered high redshift object is usually
extreme (being extreme makes its discovery easier), it is likely to
fall in the tail of the distribution. Trying to invert the argument,
and constraining cosmological parameter values after having observed a
likely extreme object is quite dangerous. This assumes
a perfect knowledge of the tail of the 
distribution from which the object is drawn, which is seldom true
on general grounds, and it is certainly not true in 
this specific case. The halo mass function differs somewhat
at the high mass end between different theoretical determinations, e.g. Jenkins et al. (2001), Press \& 
Schechter (1974), Sheth \& Tormen (1999) and numerical simulations
(see Jenkins et al. 2001).

For this reason, we do not attempt to constrain cosmological parameters
from the JKCS\,041 discovery, and we simply note that its
detection is in line with (model dependent) predictions.

\section{Summary}

We report the discovery of a massive near-infrared selected cluster of
galaxy at $z_{phot} \sim 1.9$. The evidence relies both on eleven band optical,
near-infrared and Spitzer photometry, 
and on the detection of extended X-ray emission in Chandra data.
The estimate of the redshift is based on the observed galaxy colours and 
fitting with SED of old galaxies.

The cluster is centred at $RA = 02:26:44$ and $DEC = -04:41:37$, and has
a bolometric X-ray luminosity within $R_{500}$ of $(7.4\pm1.7)\times
10^{44}$ erg s$^{-1}$. Spatial and spectral analysis indicate an X-ray
core radius of $36.6^{+8.3}_{-7.6}$ arcsec (about 300 kpc), an X-ray
temperature of $7.6^{+5.3}_{-3.3}$ keV, and a mass of
$M_{500}=2.9^{+3.8}_{-2.4}\times 10^{14}M_\odot$, the latter derived under
the usual (and strong) assumptions.

The cluster is originally discovered using a modified red-sequence
method based on near-infrared photometry, and is subsequently detected 
both by removing galaxies with 
SEDs similar to any two of the $10^4$ VVDS galaxies with $z<1.6$ in the region, 
and by using a SED fitting technique to isolate $z=1.9$ Grasil ellipticals. 
By means of the latter we find the cluster redshift to be
$1.84<z<2.12$ at 68 \% confidence.
The X-ray
detection from follow-up Chandra observations, together with statistical
arguments, discard the hypothesis of a blend of groups or a filament 
along the
line of sight. The absence of a strong radio source makes scattering of
CMB photons at X-ray energies also an unlikely explanation for the
X-ray emission.

Therefore, we conclude that JKCS\,041 is a cluster of galaxies at $z_{phot}
\sim 1.9$  with a deep
potential well, making it the highest redshift cluster currently known,
with extended X-ray emission. 

X-ray scaling relations of JKCS\,041, and other clusters at lower redshift,
will be discussed in Andreon et al. (in preparation). Sunyaev-Zeldovich
observations of JKCS\,041 are in progress at the SZ array.

\begin{acknowledgements}
We would like to thank the referee, Harald Ebeling, for a detailed
report which allowed us to improve our paper.
SA thanks Marcella Longhetti and Emanuela Pompei for useful discussions. 
Most of the Bayesian analysis in this paper benefitted from
discussions (mostly by e-mail) with 
Giulio D'Agostini, Steve Gull, Merrilee Hurn and Roberto Trotta.
This paper is based on observations obtained by UKIDSS (see standard acknowledgement
at the URL http://www.ukidss.org/archive/archive.html)
Chandra (ObsID 9368) and ESO (277.A-5028). 
BJM was partially supported during this work by NASA through Chandra guest
observer grant GO8-9117X.
JK acknowledges financial support from \emph{Deutsche Forschungsgemeinschaft}
(DFG) grant SFB 439.
\end{acknowledgements}

\end{document}